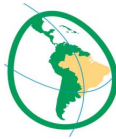

# In vitro evaluation of the effect of Ceftiofur Sodium and of a new Gentamycin Sulfate formulation on the viability of Marek disease virus

## Evaluación in vitro del efecto de Ceftiofur Sódico y de una nueva formulación de Sulfato de Gentamicina sobre la viabilidad del virus de la enfermedad de Marek

JL Chacón[1x], M Pimentel[2], A Pedroso[1], AJP Ferreira[1], D Martinez[2], C Ruelas[2]

[1]LABOR, Departamento de Patologia, FMVZ – USP. São Paulo, Brasil.
[2]CKM Salud Animal SAC. Lima, Perú.

**Abstract**
The present study evaluated *In vitro* effect of gentamicin sulfate and ceftiofur sodium on the viability of the Marek's disease virus. The titer of cell associated turkey herpesvirus (HVT) vaccine was not appreciably reduced when incubated with 50 mg/ml of gentamicin sulfate or ceftiofur sodium. Statistic difference was not found between the number of plaque-forming units (PFU) of reconstituted vaccine associated with both antibiotics 0, 15, 30 and 60 minutes after reconstitution of vaccine. The antibiotics did not considerably alter the pH values. There was a significative decrease of the titer of all vaccinal solutions when they were inoculated 30 and 60 minutes after the reconstitution of the vaccine. Nevertheless, these titers are higher than the required titers to protectect against the Marek disease.

### Introduccion

La enfermedad de Marek (EM) es una enfermedad neoplásica de las aves causada por el virus altamente contagioso de la enfermedad de Marek (VEM). La inmunización usando vacunas vivas representa una medida primordial en los programas de control. El uso masivo de las vacunas en las últimas tres décadas ha significado una reducción de la EM (Nair, 2004).Diferentes tipos de vacuna son utilizadas, tanto individualmente o en combinación. A pesar que las vacunas a virus libre y asociadas a células están disponibles, estas últimas son más ampliamente utilizadas debido a su mayor efectividad en la presencia de anticuerpos maternales (Witter & Schat, 2003). La eficacia de la vacuna está relacionada a la integridad de las células que contienen las partículas virales (Landman & Verschren, 2003). En las incubadoras, es común la administración de la vacuna de Marek junto con un antibiótico para la prevención de infecciones en pollitos de 1 día de edad. El objetivo de este trabajo fue evaluar el efecto *In vitro* de dos antibióticos inyectables, uno de ellos teniendo como principio activo Sulfato de Gentamicina, y el otro Ceftiofur sódico, sobre la viabilidad del virus vacunal asociado a células, cepa HVT de la enfermedad de Marek.

### Material y Métodos
**Antibióticos**
En este experimento fueron utilizados dos antibióticos, el primero conteniendo Sulfato de gentamicina (50mg/ml) (GENTANHEL I-5%®, CKM Salud Animal) como principio activo, y el segundo, conteniendo Ceftiofur sódico (50mg/ml) (EXCENEL®, Pfizer).

**Cultivo celular**
Cultivo primario de fibroblastos fue preparado a partir de embriones de gallina libres de patógenos específicos (SPF) de 10 días de edad. Las células fueron sembradas en placas de cultivo de células, 48 horas antes de la inoculación.

**Reconstitución de la vacuna**
Fue utilizada la vacuna congelada CRYOMAREK (Merial, Brasil) conteniendo virus vivo de la enfermedad de Marek (cepa HVT). La vacuna fue resuspendida de acuerdo a las indicaciones del fabricante en diluyente comercial de 200 ml. En seguida, 4 ml del producto GENTANHEL I-5%® y 1.6 ml de EXCENEL® (diluido según las especificaciones del fabricante) fueron diluidas en 196 y 198.4 ml de la vacuna reconstituida respectivamente.
Fueron realizadas diluciones seriadas de 1:4, 1:16, 1:64, 1:256, 1:1024 y 1:4096 con diluyente comercial en la siguientes soluciones: (1) vacuna reconstituida con diluyente comercial, (2) vacuna reconstituida con diluyente comercial y conteniendo GENTANHEL-I-5%®, (3) vacuna reconstituida con diluyente comercial y conteniendo EXCENEL® y (4) control negativo. Como control negativo, fue utilizado medio de mantenimiento de células 209 (Laboratorio Biovet, Brasil). El pH de las cuatro soluciones fueron medidas a través de pHmetro (UB10 Denver Instrument).

### Inoculación

Las cuatro soluciones fueron mantenidas a temperatura ambiente hasta el momento de ser inoculadas en las placas de cultivo celular. Cada solución comprendió de 12 repeticiones y fueron inoculadas 0, 15, 30 y 60 minutos después de preparada la solución vacunal. Las placas de cultivo de células inoculadas fueron incubadas en estufa de a 37°C con 5% de $CO_2$ durante 6 días. Una hora después de la inoculación, las soluciones vacunales fueron substituidas por medio de mantenimiento.

### Conteo de las Unidades Formadoras de Placa

Con microscopio óptico invertido (Leica® MPS 60) fue realizado el conteo de la unidades formadoras de placa (PFU) en las 12 repeticiones de las dos últimas diluciones de cada inóculo y en los cuatro tiempos establecidos.

### Análisis estadístico

Para el análisis de los datos fue utilizado el programa estadístico Graph Pad Instat 3.0 siendo los promedios comparados por el test Tukey-Kramer a 5% de significancia.

### Resultados e Discussion

Al sexto día post-inoculación fue realizado el conteo de PFU. El recuento fue realizado a partir de las dos últimas diluciones (1:1096 y 1:4096). Fue posible observar las placas características del virus de Marek en los controles positivos (**Foto 1**), y en los controles negativos fue observada la monocamada de células intacta. En el caso de las células inoculadas con la vacuna reconstituida conteniendo GENTANHEL I-5%® y EXCENEL®, fueron observadas placas idénticas a las observadas en los controles positivos, los cuales consisten en agrupamientos de células redondas y degeneradas.

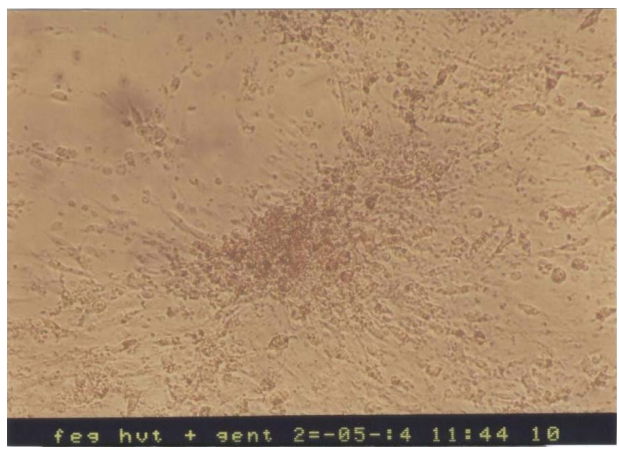

**Foto 1** - Monocapa de fibroblastos mostrando placa 6 días después de la inoculación del virus de Marek asociado a GENTANHEL I-5%. (100X).

No fue hallada diferencia estadística significativa (P>0,05) en el número de PFU del virus de Marek en las soluciones vacunales conteniendo GENTANHEL I-5%® y EXCENEL® cuando fueron comparadas con el control positivo (solución vacunal de la cepa HVT), en cada tiempo (0, 15, 30 y 60 minutos). No se encontró

**Cuadro 1** - Número de unidades formadores de placas en cultivo de fibroblastos.

| Tiempo de Inoculación | Solución Vacunal | Promedio(PFU) |
|---|---|---|
| 0 minutos | Vacuna HVT | 8448[a] |
| | Vacuna HVT + Gentanhel I-5% | 7765[a] |
| | Vacuna HVT + Excenel® | 7936[a] |
| | Control Negativo | 0 |
| 15 minutos | Vacuna HVT | 8448[a] |
| | Vacuna HVT + Gentanhel I-5% | 7765[a] |
| | Vacuna HVT + Excenel® | 7936[a] |
| | Control Negativo | 0 |
| 30 minutos | Vacuna HVT | 6400[b] |
| | Vacuna HVT + Gentanhel I-5% | 6144[b] |
| | Vacuna HVT + Excenel® | 6272[b] |
| | Control Negativo | 0 |
| 60 minutos | Vacuna HVT | 6144[b] |
| | Vacuna HVT + Gentanhel I-5% | 6059[b] |
| | Vacuna HVT + Excenel® | 5717[b] |
| | Control Negativo | 0 |

**Cuadro 2** - Valores de pH de las diferentes soluciones vacunales conteniendo el antibiótico gentamicina (Gentanhel I 5%) y Ceftiofur sódico (Excenel®).

| Solución | pH |
|---|---|
| Gentanhel-I-5% | 6.28 |
| Excenel® (diluido en diluyente comercial) | 6.81 |
| Diluyente comercial | 6.85 |
| Vacuna reconstituida | 6.89 |
| Vacuna reconstituida más Gentanhel-I-5% | 6.66 |
| Vacuna reconstituida más Excenel® | 6.88 |

diferencia estadística significativa en el título vacunal de las soluciones inoculadas a los 0 y 15 minutos después de reconstituida la vacuna; pero si hubo diferencia cuando estos títulos fueron comparados a los obtenidos por las soluciones inoculadas 30 y 60 minutos después de preparada la vacuna. Sin embargo, en todos los tiempos, el título vacunal fue superior al requerido para garantizar protección a las aves. El pH de GENTANHEL I-5%® y EXCENEL®, utilizada en proporción arriba descrita, aparentemente no alteró el pH de la solución vacunal (3.34% y 0.15% respectivamente), cuando fue comparado con el control positivo. Los hallazgos de este estudio son semejantes a los obtenidos por Eidson *et al.,* (1978), quienes no observaron disminución en el título viral en vacunas cepa HVT asociadas a células cuando fueron incubadas con 0.1, 0.2 e 0.3 mg de sulfato de gentamicina hasta por dos horas.

## Conclusion

De acuerdo con las condiciones descritas en este experimento no hay diferencia estadística significativa en el número de PFUs de la vacuna de Marek reconstituida cuando fue comparada con la vacuna asociada a GENTANHEL I-5%® y EXCENEL®, demostrando que los antibióticos no alteran significativamente la viabilidad del virus HVT.

## Referências Bibliográficas